\documentclass[review]{elsarticle}

\usepackage{lineno,hyperref}

\usepackage{amsmath}
\modulolinenumbers[5]

\journal{Journal of \LaTeX\ Templates}

\begin{document}

\begin{frontmatter}

\title{Proper scaling for triangular aperture in OAM measurement}



\author[]{Dina Grace C. Banguilan\corref{cor1} }
\ead{banguilandinagrace@gmail.com}
\author[]{Nathaniel Hermosa}
 
\cortext[cor1]{Corresponding author:}
\address{National Institute of Physics, University of the Philippines Diliman, 1101 Philippines}





\begin{abstract}
A standard triangular aperture for measuring the orbital angular momentum (OAM) of light by diffraction usually has a fixed and limited radius $R$. This possesses a crucial issue since for an increasing topological charge $m$ of an OAM beam, the radius $r$ of the beam also increases. Here, we prove experimentally our supposition. We use a dynamic triangular aperture that can be programmed to have different characteristic $R$ to diffract beams of various OAM values. By analysing the diffraction patterns with 2d-correlation, we find a minimum bound for $R$.  For a constant initial waist $w$ in the spatial light modulator and a constant position $z$ of the aperture system, we find that the radius of the aperture is unique for each $m$ value. Interestingly, this $R$ scales according to the literature-reported beam's $rms$ radius. We also show that with larger aperture, a smearing effect can be seen in the diffraction patterns which becomes a setback on discerning fringes for the measurement of the topological charge value and thus of the OAM of light. Employing this limit is required if a precise measurement of the OAM of light is needed.
\end{abstract}

\begin{keyword}
Optical modes, beam propagation, diffraction, beam shaping
\MSC[2010] 00-01\sep  99-00
\end{keyword}

\end{frontmatter}


\section{Introduction}

In 1992, orbital angular momentum(OAM) of light has been introduced. It has a well-defined value of $m\hbar$ per photon, where $m$ is an unbounded integer number referred to as the topological charge (TC) of the mode \cite{Allen}. Present in the transverse electric field of beams carrying this information is a phase factor $e^{im\phi}$. This additional factor causes the twisting of the phase fronts and thus forms an ill-defined phase or phase singularity called optical vortex (OV) whose magnitude changes over each $2\pi$ multiple of $m$  \cite{Nye}. 
The presence of the phase factor $e^{im\phi}$ implies that the output beam has an $m$-charged OV nested inside. Thus imposing such phase structure onto a laser beam creates an OAM mode. One conceptually easy way to generate an OAM mode is to pass a plane wave through a transparent material called spiral phase plate(SPP) \cite{Massari}. Since the height corresponds to a phase difference of $2\pi m$, the device imprints an azimuthal phase profile of $e^{im\phi}$ on an incoming wave. Holographic methods for creating OAM also exist, which are closely related to the principle of SPP. Generation of optical singularities can be made by computer-generated holograms(CGHs) which can have the form of a spiral Fresnel zone plate \cite{Heckenberg} or of a pitched-forked pattern \cite{Mirhosseini}. The idea is to embed the phase shift to the incident beam. The pattern can either be printed or loaded onto a microelectromechanical system,i.e. spatial light modulators (SLMs) and digital micromirror devices (DMDs), where the phase or amplitude can be modulated easily, see for example in Refs. \citep{Zambale,Reicherter}.
Over the past two decades, various applications of singular beams have been reported. The fact that photons can carry infinite OAM modes offers potential to increase the capacity of telecommunication systems \cite{Gibson}. In biology, it has been reported that rates of specific optical processes depend on the angular momentum of light of both spin and orbital components \cite{Forbes}. OAM beams are also used in optical tweezers where the intrinsic and extrinsic angular momentum of light can be transferred to microparticles and even to living cells causing a spin or a rotation that is significant in understanding different mechanisms on various types of particles and for micromanipulation \cite{Padgett2,Padgett3}.

Efficient characterization of OAM beams is therefore needed for the realization of these applications. One of the successful methods involves the rich relationship between OAM beams and diffraction phenomena. Depending on the OAM of the incident beam, the structured diffraction patterns form into optical lattices which turns out to be related to the TC value. For example, traditional single-slit \cite{Ferreira,Ghai} and double-slit \cite{Sztul} diffraction problems can be used to analyze the dark and bright bands of light giving rise to the determination of the OAM of light. Interference methods have also been reported for a similar purpose. For example, a group of researchers from Glasgow have efficiently sorted different OAM states of single photons through channels of Mach-Zehnder interferometer \cite{Leach}, and Lv's group reconfigured this technique by the insertion of Dove prism to measure higher order TC of the beam through interference intensity analysis \cite{Lv}. Besides these, schemes involving shaped apertures have been also proposed to sort and detect OAM states. The group of Hickmann pioneered the use of triangular aperture to investigate the vortex phase structure of the beam from the lattice properties of the diffraction patterns formed \cite{Hickmann}. Methods based on multipoint-interferometer \cite{Berkhout,Liu_triangularmultipoint} and polygonally-shaped slits and apertures \cite{Li_dualtriangle,Hickmann2,Liu_diamond,Liu_hexagon,Acevedo} have been engineered to demonstrate the relationship between the generated diffraction patterns and the amount of OAM carried by light. All these stated techniques turn out to provide information about the TC of the beam which characterizes the incident light field. Although the determination of OAM for beams is well-established, the rich features of it has remained a great challenge to unveil. 

However, OAM beams have more complex structures and inherent features that can limit its characterization.  Spatial dispersion, for example, has been reported to occur naturally in generating broadband vortices which poses problem in determining the TC value especially for higher order beams \cite{Anderson}. This feature causes smearing of the individual spots near the edges of the pattern making it impossible to count the number of spots and so limiting the applicability of the diffraction technique. There are also types of beams called elliptical vortex beams. To investigate such beams, one must consider the elliptical symmetry of the beam to design the most appropriate aperture shape to be used to diffract the beam. Melo et. al. have developed an isosceles triangular aperture to maximize the visibility of the diffraction features that contain the information about the topological charge as a function of the relative size of the beam \cite{Melo}. Moreover, it is known that artifacts are present in the side limit of the diffraction patterns aside from the primary maxima spots \cite{Stahl}, and this additional trait might lead to difficulty in counting the necessary lattice points and in discriminating the OAM state \cite{Hickmann3}. From these, we recognize the necessity of an optimal parameter or an scaling factor to observe distinct diffraction spots for an accurate OAM detection.

Here, we show that for the diffraction technique to work in obtaining the TC of the beam, one needs to consider the variation on the beam's radius and consequently on the dependence of the diffracted field on the aperture's size. In section 2, we present a theoretical investigation on the variation of the beam's radius as it propagates and we review how the TC is determined in simulations. This is followed by a demonstration of our method using a practical optical setup involving a dynamic triangular aperture. We discuss in section 4 that to design the most appropriate triangular aperture for the diffraction task, the radius of the aperture must scale with the radius of the impinging OAM beam. This aperture radius is unique for each TC value. Which means the radius of the aperture also is dependent on the TC of the beam. For apertures out of the minimum bound provided, the generated patterns are not well-defined and we cannot precisely discern the number of fringes to be counted for OAM determination. Finally, we summarize this work in section 5.

\section{Numerical Simulations and Results}

\begin{figure}[t]
\centering
\includegraphics[width=\linewidth]{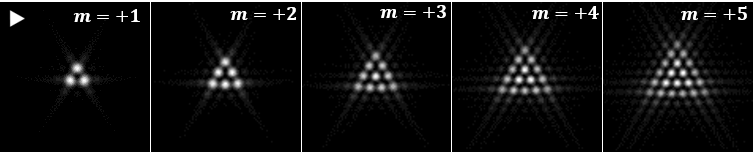}
\caption{Numerical diffraction patterns generated for $m$ ranging from one to five. The light intensity increases from black to white. The inset in the $m=+1$ shows the orientation of the triangular aperture.}
\label{fig:nresults}
\end{figure}

We start investigating the field by numerically evaluating equation \ref{eqn:field} with a triangular aperture (of constant size) using a Laguerre-Gauss(LG) beam.  
\begin{equation}
\label{eqn:field}
E_d (k_\perp)=\int\limits_{-\infty}^{+\infty} \tau(r_\perp) u_{o}(r_\perp) e^{-ik_\perp .r_\perp } \ d r_\perp
\end{equation}

To obtain the diffracted field in this form, we implement a custom code based on a commercial algorithm for fast Fourier transform(FFT) calculation. Which means that the diffracted field is just  the Fourier transform of the product of the aperture function and the incident field $u_{o}$, where the transverse vector $r_\perp$ represents the coordinates in the reciprocal space. For a specific case, the $\tau(r_\perp)$ describes the equilateral triangular aperture which has unity transmission inside the aperture and zero outside the aperture. And the far-field distribution is recorded by taking the square of the absolute field. 

However, as pointed out by Ricci et al., we note that the intensity of the propagating field has two important forms \citep{Ricci}. One comes from the theoretically expected pure $p=0$ LG beam given by

\begin{equation}
\label{eqn:LG}
u_{LG}\propto  \Bigg(\frac{r}{w}\Bigg)^{\vert m\vert}e^{\big(\frac{-r^2}{w^2}\big)} e^{im\phi}
\end{equation}

\noindent with radius of maximum amplitude equal to \cite{Padgett5}

\begin{equation}
r_{max}(m)=\sqrt{\frac{m}{2}}w
\label{RbeamLG}
\end{equation}

\noindent But the propagation of the field usually generated in the experiment results to Kummer function expressed as

\begin{equation}
\label{eqn:KB}
u_{KB}\propto   e^{\big(\frac{-r^2}{2w^2}\big)} \Bigg(\frac{r}{w}\Bigg)\Bigg[I_{(m-1)/2}\Bigg(\frac{-r^2}{2w^2}\Bigg)-I_{(m+1)/2}\Bigg(\frac{-r^2}{2w^2}\Bigg)\Bigg]e^{im\phi}
\end{equation}

\noindent where $I_{(m\pm1)/2}$ is the $m^{th}$ order Bessel function of the first kind \cite{Arfken}. This beam has a radius given by the standard deviation of the intensity distribution of the beam \cite{Divergence_Padgett}, 

\begin{equation}
        r_{rms}(m)=\sqrt{\frac{m+1}{2}} w
\label{RbeamKummer}
\end{equation}

\begin{figure}[t]
\centering
\includegraphics[width=\linewidth]{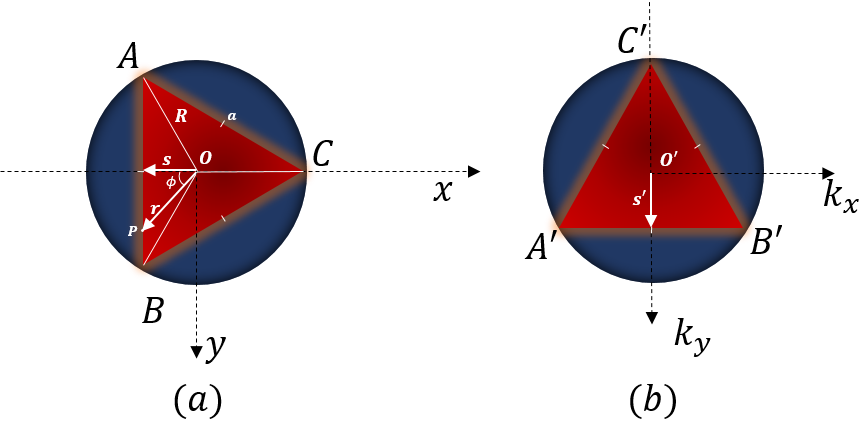}
\caption{ a) Geometry of an equilateral triangular aperture inscribed in a circle matching the global maxima of the intensity pattern of a LG beam. b) Geometry of the intensity distribution at the focal plane.}
\label{fig:2}
\end{figure}

Compared to the LG model, Kummer model of the diffracted beams has been found to agree more with the actual observations \cite{Bekshaev}. That is, distinctions between these become more significant with increasing $\vert m \vert$. Compared to the distribution of the theoretical LG beam, Kummer beams’ intensity distribution in the far-field is wider and has lower maximal value. Supposing then that the optical field ($u_{o}$=$u_{KB}$) falls on-axis on the triangle aperture as in fig. \ref{fig:2}a, and by performing eq. \ref{eqn:field} we obtain the diffraction patterns as in fig. \ref{fig:nresults}.

We observe a formation of optical lattice points of maximum intensity which becomes richer as $m$ increases. In fact the number of interference peaks grows as the sum of an arithmetic progression  given by $\frac{(m+1)(m+2)}{2}$. We note that the triangular lattice distribution results from the interference of the geometrical and boundary waves of the aperture as was recently asserted in Ref. \cite{Narag}. Basically, each aperture edge produces a light fan or ray in a direction perpendicular to its spatial orientation. Interference between the three light fans occurs only at the central region of the triangular aperture for an incident light with $m=0$, thus creating there only a single bright spot \cite{Hickmann4}. On the other hand, whenever an electric field with $m\neq0$ is incident on the aperture, the light fans interfere at positions other than the center of the aperture \cite{Stahl,Hickmann3}. Superposition of many light fans from the edges of a finite-size aperture will create parallel and equally spaced satellites forming bright spots or lobes in the Fourier plane. And we are able to replicate the results of Hickmann et. al. stating that the number of spots along the side of the distribution, $c$, is one greater than the OAM beam's topological charge \cite{Hickmann}. Since $c$ increases with $m$, the size of the intensity distribution becomes larger. And as depicted in fig. \ref{fig:2}, a slit along the  $y$-direction produces a pattern that is displaced in the transverse $k_y$ direction by a value which increases with the TC of the vortex beam as was reported in Ref.\cite{Yongshin}.

\section{Experiment}

\begin{figure}[t]
\centering
\includegraphics[width=\linewidth]{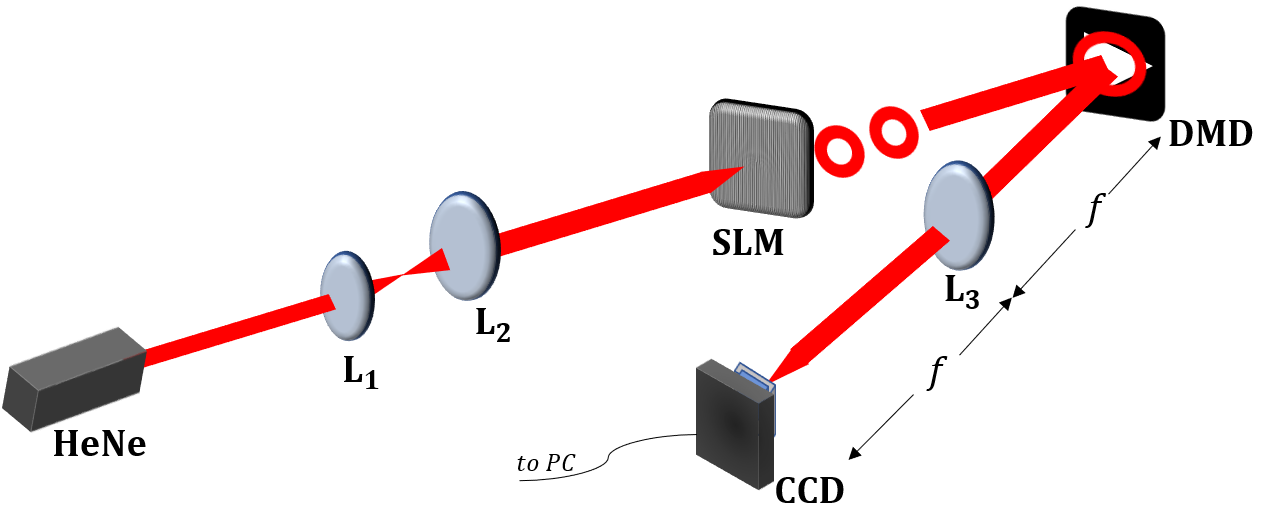}
\caption{Optical layout of the experimental setup for generating various OAM modes, for their diffraction and characterization.}
\label{fig:setup}
\end{figure}

We experimentally prove our initial results with the setup schematically shown in figure \ref{fig:setup}. The output beam of a HeNe laser operating at $633 nm$ wavelength is collimated to generate a plane wave and expanded using a system of two lenses, $L_1$ with focal length $f_1=100mm$ and $L_2$ with focal length $f_2=200mm$. The collimated beam with a constant waist $w$ illuminated the display of the spatial light modulator(SLM) to produce modes with different $m$. We limit ourselves with OAM modes with $m=1,2,3,4,5$ and $p=0$. The input of the SLM display is composed of a phase function and a linear phase mask. From the set of different orders generated from the SLM, the first diffracted order is spatially separated to impinge the digital micromirror device(DMD) where the triangular aperture is loaded. A DMD is a microelectromechanical system composed of an array of tiny mirrors. Each mirror corresponds to one pixel of the projected image and has a dimension $10.8\mu m$ $\times$ $10.8\mu m$.  Note that the uploaded apertures are scaled to correct the structural shape produced by the device \cite{Zambale2}. A $2f$ lens system given by $L_3(f_3=100mm)$ is implemented after the DMD to carry out the optical Fourier transform of the field in the aperture plane onto the CCD detection plane. This is the physical realization of the integration in eq. \ref{eqn:field}. The transverse intensity patterns corresponding to the Fraunhofer diffraction are registered in the CCD camera.  The SLM and the DMD are controlled using a computer. 

\section{Experimental Results and Discussion}

\begin{figure}[t]
\centering
\includegraphics[width=\linewidth]{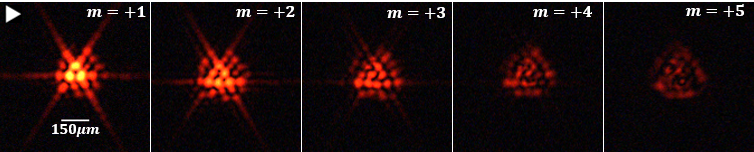}
\caption{Experimental diffraction patterns for $m$
 ranging from one to five. The light intensity increases from black to yellow-orange. The inset in the $m=+1$ shows the orientation of the triangular aperture with radius $R=2.16mm$.}
\label{eresults}
\end{figure}

We see from fig. \ref{eresults} that the diffraction of an OAM beam by the triangular aperture creates a distribution containing optical lattices. Although we are able to replicate our simulations saying that $m=c-1$, the diffraction patterns produced especially when $m>1$ are not well-defined. Moreover, we observe from our simulations a little to no effect of the constant aperture size for diffracting different OAM states. But this does not hold in the experiment. From a combination of a constant aperture size and different $m$-values, diffraction patterns with different contrast and intensities are detected. We presume that this is due to the mismatch between the aperture's characteristic radius $R$ and the beam's radius $r$. Eq. \ref{RbeamLG} and \ref{RbeamKummer} clearly state that the respective radius of the theoretical and physical beam increases and scales with $m$. Here, we believe that the size of the diffracting aperture should also vary accordingly with the incident beam's size. But the question is, \textit{by how much?} 

We put forward on this problem by performing another set of experiment. We let a single OAM-state $m$ to impinge a triangular aperture of various radii $R$. Shifting from one aperture size to another is made easier by our dynamic aperture. The aperture's radius $R$ refers to the distance from the center of the triangle to one of its vertices when inscribed in a circle, as illustrated in fig. \ref{fig:2}a.

\begin{figure}[htbp]
\centering
\includegraphics[width=\linewidth]{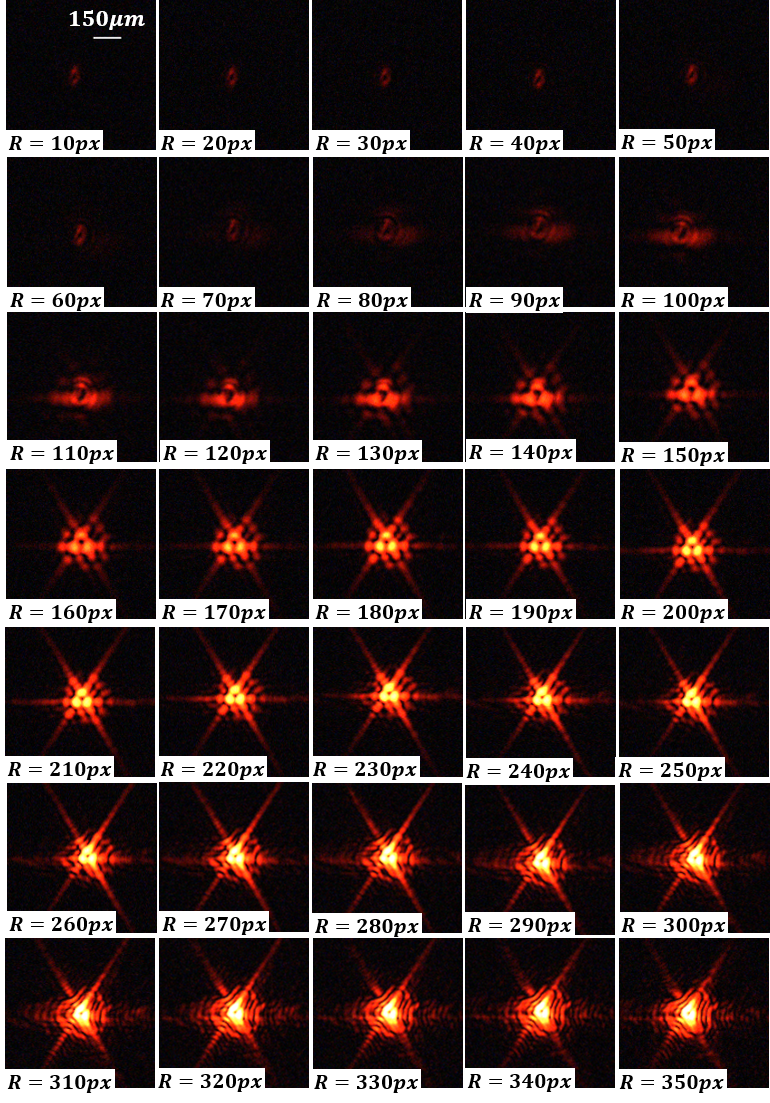}
\caption{Generated diffraction patterns with $m=1$ for various sizes of triangular aperture. ($1px=10.8\mu m$)}
\label{fig:m1sizes}
\end{figure}

Our next goal is to determine the diffraction patterns in the far-field region of a beam with OAM and see the effect of various size of aperture to the pattern. Here, we model the behavior of the aperture as follows:
    \begin{equation}
        R(m)=\alpha(m)R_{o}
    \label{Raperture}
    \end{equation}
where
    \begin{equation}
        \alpha(m)=\begin{cases}\sqrt{\frac{m+1}{2}}=\alpha_{1} \\ \sqrt{\frac{m}{2}}=\alpha_{2}\end{cases}
        \label{factor}
    \end{equation}
    
and $R_o$ is a constant parameter. 
We choose the first model $\alpha_{1}$ to represent the change in aperture's size to vary in accordance with Kummer beams' radius variation as described in eq. \ref{RbeamKummer}. The second case, $\alpha_{2}$, derives from the representative radius for the main energy corresponding to the maximum intensity of a single-ringed Laguerre-Gaussian laser mode given by eq. \ref{RbeamLG}. The result for a combination of $m=1$ and thirty five(35) apertures of different radii is shown in fig.  \ref{fig:m1sizes}. We are also able to test the dependence of the diffraction patterns for $m=2,3,4,5$ on different sizes of $R$. The results follow the similar trend.

It is noticeable that the structure of each pattern differs from one another in terms of brightness, contrast and symmetry of the triangular distribution. Particularly, the pattern is seen to be more prominent for a specific range of the aperture's radius. Relying only on these pictures, for a beam with TC equal to one and a constant OAM waist $w$, the aperture with radius ranging from $R=160px$ to $R=250px$ produces a diffraction pattern with a maximum visibility of the three lobes forming it. For values of $R$ below this bound, the patterns are indistinct and have low intensity such that counting the number of spots will be impossible. Above this limit, the patterns are more intense, but one lobe coalesces with the adjacent lobes which becomes problematic in discriminating diffraction fringes. In general, a smearing effect is seen whenever the aperture used to diffract the 1-charged beam is inappropriate, and this becomes more obvious for larger diffracting apertures. This observation means that the intensity patterns by a triangular polygon is distinct for different aperture radii and beam sizes. From these, we infer that well-defined patterns start to appear only if the diffracting object matches the radius of the incident OAM beam define by a factor related to these two parameters.  

\begin{figure}[t]
\centering
\includegraphics[width=\linewidth]{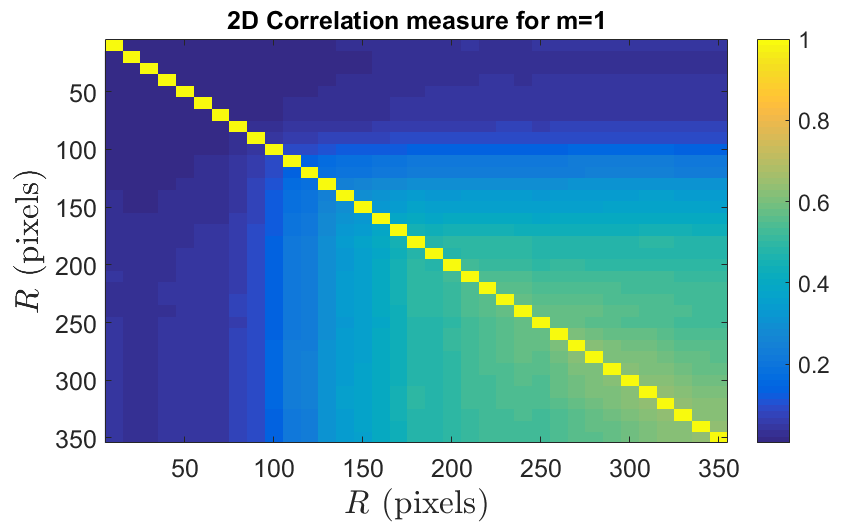}
\caption{Obtained 3D plot of the correlation among the thirty-five diffraction patterns. The mean $R$ calculated from the four trials is $R=1.78mm$.}
\label{fig:m1corr}
\end{figure}

To verify this observation, we test numerically the relationship between and among the patterns by employing a 2d-correlation method. The correlation coefficient $C$ between two images is given by eq. \ref{eqn:correlation},

\begin{equation}
\label{eqn:correlation}
C=\frac{\sum_{m}\sum_{n}(X_{mn}-\bar{X})(Y_{mn}-\bar{Y})}{\sqrt{(\sum_{m}\sum_{n}(X_{mn}-\bar{X})^2) (\sum_{m}\sum_{n}(Y_{mn}-\bar{Y})^2)}}
\end{equation}

\noindent where $\bar{X}$ and $\bar{Y}$ are the mean intensities of images $X$ and $Y$, respectively. A higher correlation indicates a better similarity between the two compared images. For example, a pattern produced with different $R$ is compared to all the other thirty-five patterns and obviously when compared to itself we find $C=1$. A 3D plot of the correlation among the patterns for $m=1$ is shown in fig. \ref{fig:m1corr}. We test our experimental results for four(4) different correlation thresholds, $C\ge.2$, $C\ge.3$, $C\ge.4$ and $C\ge.5$. These correlation values are predetermined. This range allows a general trend on the variation of $R$ for each $m$. At threshold values other than these, the variation of $R$ with respect to the TC value is unsteady and cannot be used to interpret the general result. These correlation thresholds represents the number of trials made in this study. Thus, by averaging the values provided by these, we get a minimum $R$ appropriate for diffracting different OAM states. The obtained mean $R$ from the four trials using $m=1,2,3,4,5$ are $1.78mm$, $2.11mm$, $2.48mm$, $2.75mm$ and $3.08mm$, respectively, with corresponding variances equal to $\pm.32mm$, $\pm.32mm$, $\pm.24mm$, $\pm.19mm$ and $\pm.19mm$. Referring to the 3D plot, we see that at a constant correlation threshold, we obtained different minimum bounds for the $R$ unique for each TC value. For a constant $m$, a smaller correlation threshold $C$ gives the smallest $R$ and similarly a larger $C$ allows a larger aperture size. 

Moreover, we see that for the case where $m=1$, the  minimum $R$ obtained from the 2d-correlation method is within the range of values provided by our human eye, that is, from relying solely on the pictures in fig. \ref{fig:m1sizes}. The correlation values depend also on the sensitivity set on the imaging device. Hence, the images are  obtained at the highest sensitivity level provided by our CCD camera.  With this setting, all information contained in the patterns are detected, including noise and parasitic interference.

\begin{figure}
    \centering
    \includegraphics[width=.8\textwidth]{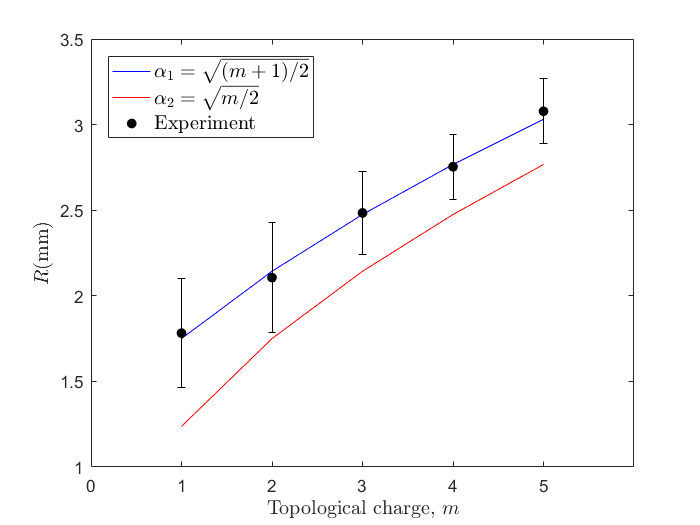}
    \caption{Plot of the radius of the aperture versus the topological charge values for the different models for the OAM beam's radius.}
    \label{fig:plot2models}
\end{figure}

We perform a linear regression to test the two models. Our calculations show a strong positive linear relationship between our experimental data set and the first model. The recorded average $R$ fits well with the first model with an \textit{$r_{P}^2$}(regression statistic) value equal to $0.9957$ which implies that this model accounts $99.57\%$ of the variance in the obtained $R$. 
The Laguerre-Gauss model also has a linear relationship with an \textit{$r_{P}^2$} equal to $0.8513$. Based on the values of $r_{P}^2$ we obtained, it is clear that the first model best describe how the aperture size should change with $m$.

We then examine the average $R$ values with the two models in eq. \ref{factor} with respect to the charge $m$. From the plot in fig. \ref{fig:plot2models}, interestingly, the curve of the experimental $R$ best fitted in the first model by a factor $R_o$. Which means that the aperture size vary accordingly with the change in the size of the beam at constant waist and propagation distance by a scale factor equal to $\alpha_1=\sqrt{\frac{m+1}{2}}$. We generalize this result by the equation 
\begin{equation}
\label{eqn:final}
R(m)=\sqrt{\frac{m+1}{2}}R_o
\end{equation}
So, for $m=1$ clearly $R(1)=R_o$. This result can be updated to obtain the radius of the aperture applicable for diffracting $m=2$ and so on. If one set $R_o=R(m=1)$ then the value for other $R's$ can be obtained. Note, however, that $R(m=1)$ or $R_o$ is chosen depending on the discretion of the observer. It is a "test" size selected to provide a maximum visibility of the features on the diffraction pattern for $m=1$ in particular. We observe also that the general trend of the obtained experimental values are also close to that of the second model. This means that the second model arises as a limiting case in the scaling of the triangular aperture for large $m$-values.

By following our developed technique, the results of this study can be directly applied to other vortex beams or vortex beams generated by other means.  

\section{Summary and Conclusion}


We presented a scheme to design the most appropriate size of the triangular aperture when employing the diffraction technique in measuring the OAM of light. Specifically, we examined the variation of aperture's radius and its effect to diffract different OAM modes. 
From our observations, we say that for a constant initial beam waist $w$ incident to the spatial light modulator and at a constant propagation distance $z$, the aperture's radius must scale with $\sqrt{\frac{m+1}{2}}$. This factor turns out to be similar to the scaling law of the radius of root-mean-square of the beam by some constant fitting parameter $R_o$. We believe that this scalability makes the detection method very useful for real optical systems where the size of OAM modes are expected to increase.




\bigskip
\noindent \textbf{Funding.} Department of Science and Technology- Philippine Council for Industry, Energy and Emerging Technology Research and Development(DOST-PCIEERD).
\vspace{.5cm}

\noindent \textbf{Acknowledgments.} N. Hermosa is a recipient of the University of the Philippines Office of the Vice-President for Academic Affairs Balik-PhD program (OVPAA-BPhD 2015-06). D.G.C. Banguilan is a scholar of the Department of Science and Technology-Science Education Institute (DOST-SEI) thru its Accelerated Science
and Technology Human Resource Development Program (ASTHRDP).

\bibliography{mybibfile}

\end{document}